\documentclass[conference]{IEEEtran}
\IEEEoverridecommandlockouts

\usepackage{blindtext}

\usepackage{etoolbox}
\usepackage{lipsum} %<---- For dummy text

\usepackage{siunitx}
\usepackage{pifont}
\usepackage{graphicx}
\usepackage{amsmath, amssymb,textcomp}
\usepackage{booktabs}
\usepackage{url}
\usepackage{subcaption} 
\usepackage{algpseudocode}
\usepackage{paralist}
\usepackage{color}
\usepackage{relsize}
\usepackage{stmaryrd}
\usepackage{epsfig} % for postscript graphics files
\usepackage{bm} % assumes amsmath package installed
\usepackage{listings,lstautogobble}
\usepackage{color}
\usepackage{mathtools}
\usepackage{mathrsfs}
\usepackage{xspace}
\usepackage{soul,xcolor}
\usepackage{stfloats}
\usepackage{verbatim}
\usepackage{epstopdf}
\usepackage{listings}
\usepackage{parcolumns}
\usepackage{color}
\usepackage{mathtools}
\usepackage{mathrsfs}
\usepackage{xspace}
\usepackage{soul,xcolor}

\usepackage{verbatim}
\usepackage[utf8]{inputenc}
\usepackage{calrsfs}
\usepackage{rotating,multirow}
\usepackage{caption}

\usepackage[most]{tcolorbox}
\usepackage{inconsolata}

\usepackage{blindtext}

\usepackage{etoolbox}
\usepackage{graphicx}
\usepackage{balance}
\usepackage{comment} 
\usepackage{tcolorbox}
\usepackage{amsthm}

\usepackage{dsfont}
\usepackage{makeidx}  % allows for indexgeneration
\usepackage{latexsym,amsfonts}
\usepackage{graphicx}
\usepackage{caption}

\usepackage{graphics} % for pdf, bitmapped graphics files
\usepackage{epsfig} % for postscript graphics files
\usepackage{mathptmx} % assumes new font selection scheme installed
\usepackage{amsmath,bm} % assumes amsmath package installed
\usepackage{amssymb}  % assumes amsmath package installed
\usepackage{listings}
\usepackage{color}

\usepackage{graphicx}
\usepackage{caption}
\DeclareCaptionType{copyrightbox}
\usepackage{upgreek}
\usepackage{siunitx}
\usepackage{varwidth}
\usepackage{tikz}
\usetikzlibrary{shapes,shadows,calc,snakes}
\usetikzlibrary{decorations.pathmorphing,patterns}
\usetikzlibrary{decorations}

\usepackage{ifthen}
\usepackage{epstopdf}
\usepgflibrary{arrows}
\usepackage{circuitikz}
\usepackage{makecell}
\usepackage[ruled,vlined,linesnumbered,noresetcount]{algorithm2e}
\usepackage{sidecap}
\usepackage{stackengine}
\def\delequal{\mathrel{:=}}
\newcommand\Mark[1]{\textsuperscript{#1}}

\newcommand{\s}{\scriptscriptstyle}

\newcommand{\mathbbm}[1]{\text{\usefont{U}{bbm}{m}{n}#1}} % from mathbbm.sty

\DeclareMathAlphabet{\mathcal}{OMS}{cmsy}{m}{n}

\DeclareMathOperator\diag{\mathsf{diag}}
\DeclareMathOperator\T{\mathsf{T}}
\DeclareMathOperator\tr{\mathsf{tr}}

\newtheorem{example}{Example}

\newtheorem{lemma}{Lemma}
\def\BibTeX{{\rm B\kern-.05em{\sc i\kern-.025em b}\kern-.08em
    T\kern-.1667em\lower.7ex\hbox{E}\kern-.125emX}}
\begin{document}

%\author{\IEEEauthorblockN{Name Surname\IEEEauthorrefmark{1},
%Name Surname\IEEEauthorrefmark{2}, Name Surname\IEEEauthorrefmark{3} and
%Name Surname\IEEEauthorrefmark{4}}
%\IEEEauthorblockA{University of California, Los Angeles\\
%Wherever\\
%Email: \IEEEauthorrefmark{1}author.one@add.on.net,
%\IEEEauthorrefmark{2}author.two@add.on.net,
%\IEEEauthorrefmark{3}author.three@add.on.net,
%\IEEEauthorrefmark{4}author.four@add.on.net}}

\title{Event-Triggered Diffusion Kalman Filters}%\vspace{-10mm}}

\author{\IEEEauthorblockN{Amr Alanwar\Mark{1},
Hazem Said\Mark{2}, Ankur Mehta\Mark{3}, and Matthias Althoff\Mark{1}}
\IEEEauthorblockA{\Mark{1}Technical University of Munich, \Mark{2}Ain Shams University, \Mark{3}University of California, Los Angeles \\
Emails: \{alanwar, althoff\}@tum.de, hazem.said@eng.asu.edu.eg, mehtank@ucla.edu }\vspace{-9mm}}

%\title{Event-Triggered Diffusion Kalman Filters\\[.75ex] 
%  {\normalfont\large 
%    Amr Alanwar\Mark{(1)}, Hazem Said\Mark{(2)}, Ankur Mehta\Mark{(3)}, Matthias Althoff\Mark{(1)}%
%  }\\[-1.5ex]
%}

%\author{
%    \IEEEauthorblockA{%
%        \Mark{(1)}Technical University of Munich
%    }
%    \and
%    \IEEEauthorblockA{%
%        \Mark{(2)}Ain Shams University
%    }
%    \and
%    \IEEEauthorblockA{%
%        \Mark{(3)}University of California, Los Angeles
%    }
%\vspace{-12mm}}

\maketitle

\begin{abstract}

%On Event-Based Distributed Kalman Filter With Information Matrix Triggers

%The performance of a distributed network state estimation problem depends strongly on collaborative signal processing, which often involves excessive communication and computation overheads on a resource-constrained sensor node. 
%In this work, we approach the distributed estimation problem from the viewpoint of sensor networks to design a more efficient algorithm with reduced overheads, while still achieving the required performance bounds on the results.
Distributed state estimation strongly depends on collaborative signal processing, which often requires excessive communication and computation to be executed on resource-constrained sensor nodes. To address this problem, we propose an event-triggered diffusion Kalman filter, which collects measurements and exchanges messages between nodes based on a local signal indicating the estimation error. On this basis, we develop an energy-aware state estimation algorithm that regulates the resource consumption in wireless networks and ensures the effectiveness of every consumed resource. The proposed algorithm does not require the nodes to share its local covariance matrices, and thereby allows considerably reducing the number of transmission messages. To confirm its efficiency, we apply the proposed algorithm to the distributed simultaneous localization and time synchronization problem and evaluate it on a physical testbed of a mobile quadrotor node and stationary custom ultra-wideband wireless devices. The obtained experimental results indicate that the proposed algorithm allows saving $86\%$ of the communication overhead associated with the original diffusion Kalman filter while causing deterioration of performance by $16\%$ only. We make the Matlab code and the real testing data available online\footnote{https://github.com/aalanwar/Event-Triggered-Diffusion-Kalman-Filters}.

\end{abstract}

\begin{IEEEkeywords}
Event-triggering, diffusion Kalman filter, localization, time synchronization.
\end{IEEEkeywords}

   % WSN and DKF
   %Wireless sensor networks consist of spatially distributed sensor devices %which play a vital role to monitor the state of network 
   %State estimation algorithms across wireless sensor networks offer many advantages and services in emergency rescue, homeland security, military operations, habitat monitoring, and home automation services \cite{conf:wirelessnetwork}. %Such critical services would require maintaining some guarantees on 
   %Besides ensuring the accuracy of the state estimation, one has to consider power constraints \cite{conf:powerconstraint}, limitations in terms of bandwidth \cite{conf:bwconstraint}, and limitations in computation and communication \cite{conf:compconstraint, conf:comoverhead}. %Thus, many design challenges are imposed in the state estimation algorithms, which get much attention in recent years among the researchers \cite{conf:estchallenge}. Also, due to the importance of resource awareness in a broad range of emerging wireless network algorithms running on resource-constrained commodity platforms, it is imperative to rethink how such resources are handled in the state estimation algorithms across the sensor networks. And what is the trade-off between system performance and the consumed resources in the wireless sensor network?

\section{Introduction}
   % event trigger
   The State estimation algorithms used in wireless sensor networks enable services in various fields, such as emergency rescue, homeland security, military operations, habitat monitoring, and home automation services \cite{conf:wirelessnetwork}. In addition to ensuring the accuracy of the state estimation, one has to consider power constraints \cite{conf:powerconstraint}, limitations in terms of bandwidth \cite{conf:bwconstraint}, and limitations in computation \cite{conf:compconstraint} and communication \cite{conf:comoverhead}. One of the most widely applied estimation algorithms for sensor networks is the distributed Kalman filtering algorithm \cite{conf:avgconsensus3}. Among distributed Kalman filters, diffusion algorithms \cite{conf:diffusion-org} have favorable properties with respect to performance and robustness in terms of handling node and link failures.  %however, it requires a tremendous amount of message exchange and computation overhead \cite{conf:diffusion}. 

   %State estimation algorithms across wireless sensor networks offer many advantages and services in emergency rescue, homeland security, military operations, habitat monitoring, and home automation services \cite{conf:wirelessnetwork}. Besides ensuring the accuracy of the state estimation, one has to consider power constraints \cite{conf:powerconstraint}, limitations in terms of bandwidth \cite{conf:bwconstraint}, and limitations in computation \cite{conf:compconstraint} and communication \cite{conf:comoverhead}. One of the most popular estimation algorithms for sensor networks is the distributed Kalman filtering algorithm \cite{conf:avgconsensus3}. Among distributed Kalman filters, diffusion algorithms \cite{conf:diffusion-org} have favorable properties with respect to performance and robustness of node and link failures.  %however, it requires a tremendous amount of message exchange and computation overhead \cite{conf:diffusion}. 

 %We investigate our algorithm capabilities, which restricts the amount of processing, sensing, and communication. 

The performance of the diffusion Kalman filters \cite{conf:diffusion-org} depends on frequent measurements and message exchange between nodes. However, the capabilities of individual nodes are limited, and each node is often battery-powered. Therefore, decreasing communication overhead and the number of measurements is of great importance. The main open question to be considered is not how great estimation an algorithm could achieve, but rather to what extent it is capable of satisfying the application needs while saving resources.
% There is no meaning of spending more resources, while the application need is much less.
To address this question, we propose an event-triggered diffusion Kalman filter that restricts the amount of processing, sensing, and communication based on a local signal indicating an estimation error without requiring the nodes to share their local covariance matrices. Therefore, the number of transmission messages compared to the nominal distributed diffusion Kalman filter algorithm is significantly reduced. In particular, we characterize the trade-off between the consumed resources and the corresponding estimation performance. 

   As a representative application of distributed state estimation, we consider localization and time synchronization. %With the growing prevalence of wireless devices, it is important to coordinate timing and location among Internet-of-Things (IoT) devices. 
   Maintaining a shared notion of the time is critical for ensuring acceptable performance and robustness of many cyber-physical systems (CPS). % like swarm robotics \cite{conf:formationctrl} Furthermore, position estimation is necessary for different fields. %such as military \cite{conf:military}, indoor and outdoor localization \cite{conf:indoor}, security surveillance, and wildlife habitat monitoring.
   Furthermore, position estimation is a crucial task in different fields. Nevertheless, localization and time synchronization algorithms require a significant amount of collaboration efforts between individual sensors, and therefore, the proposed approach is particularly helpful for this application. More specifically, we apply the proposed event-triggered diffusion Kalman filter on D-SLATS \cite{conf:d-slats,conf:d-slats-cdc}, which is a distributed simultaneous localization and time synchronization framework. 
 % This collaborative signal processing nature of sensor networks requires significant research efforts for energy management. For example, just the decision of whether to do the collaborative message exchanges or some local processing has a significant implication on the overall energy and lifetime.

   % contribution
In the present study, we make the following contributions:
\begin{itemize}
\item Introducing the event-triggered distributed diffusion Kalman filter to reduce communication, computational, and sensing overheads.
\item Showing that our event-triggered estimator is unbiased and deriving the relationship between the triggering signal and the expected error covariance.
\item Applying the proposed strategy in localizing and time-synchronizing of distributed nodes in an ad-hoc network. 
%\item Proving the stability of the event trigger D-SLATS \cite{conf:d-slats}.
\item Evaluating the proposed strategy on a real testbed using custom ultra-wideband wireless devices and a quadrotor.%, thereby representing a network of both static and mobile nodes.
\end{itemize}

   % organization
   The rest of the paper is organized as follows. Section \ref{sec:related} provides an overview on the related work. Section \ref{sec:mot} gives the motivation behind our chosen triggering condition. Then, we present the proposed algorithm and corresponding theoretical analysis in Sections \ref{sec:algo} and \ref{sec:analysis}, respectively. Section \ref{sec:eval} illustrates the application to localization and time synchronization, provides the description of the experimental setup, and evaluates the proposed algorithm on static and mobile networks of nodes. Finally, Section \ref{sec:conc} summarizes some concluding and discussion remarks.
   
   %Section \ref{sec:system} provides an overall overview of the system model under our study. 

\section{Related Work} \label{sec:related}
%diffusion LMS, and diffusion RLS  Diffusion Kalman filtering and smoothing algorithms 
%The related work can be categorized as follows:
We first discuss the general state estimation algorithms followed by centralized and distributed event-triggered estimators.
\subsection{State Estimation Algorithms}
%Under this umbrella, we have centralized and distributed estimation algorithms. We are more interested in the distributed algorithms due to their advantages over the centralized ones. 
%Algorithms for diffusion least-mean squares \cite{conf:difflms}, diffusion recursive least-squares \cite{conf:diffrls}, and diffusion Kalman filtering \cite{conf:diffkalman} have been proposed. 
%Distributed estimation algorithms are widely used in wireless networks to reconstruct the system state from noisy measurements. 

Estimation algorithms based on average consensus have been analyzed in \cite{conf:avgconsensus3,conf:avgconsensus1,conf:avgconsensus2}. The distributed estimation algorithm proposed in \cite{conf:avgconsensus4} is intended for extremely large-scale systems. The main idea is to approximate the inverse of the large covariance matrix $P$ by using the L-banded inverse and the distributed iterate collapse inversion overrelaxation (DICI-OR) method \cite{conf:dici-or}; however, it requires a large amount of computation resources. Due to limited resources available in wireless sensor networks, many investigations have been made seeking to decrease the communication and computation overheads, while preserving the acceptable performance. This has served as the basis for the studies in the following categories.
%These algorithms lack of the resource-aware aspect, which is an essential need in the wireless network.

\subsection{Centralized Event-Triggered Estimation Algorithms}

The event-triggered scheme has already been applied to network estimation problems. It was first proposed with regard to the centralized estimation problems. The send-on-delta method is proposed for Kalman filters in \cite{conf:kfsendondelta}, where the sensor data values are transmitted only upon encountering a user-defined change. An event-triggered sensor data scheduler based on the minimum mean-squared error (MMSE) has been proposed in  \cite{conf:shedulingwu2013event}. In \cite{conf:kfcovevent}, the variance-based triggering scheme has been introduced, implying that each node runs a copy of the Kalman filter and transmits its measurement only if the associated measurement prediction variance exceeds a chosen threshold. The properties of the set-valued Kalman filters with multiple sensor measurements have been analyzed in \cite{conf:kfcovevent2}. Open-loop and closed-loop stochastic event-triggered sensor schedules for remote state estimation have been proposed \cite{conf:yilin}.

In general, the required amount of communication can be reduced by using an event-triggered scheme in which the sensor and estimator are not at the same node, as discussed in \cite{conf:eventACC,conf:li2013polynomial}. Moreover, a discrete-time approach is proposed in \cite{conf:groff2016observer} to address the same concern. The importance of considering the effects of external disturbances and measurement noise in the analysis of event-triggered control systems is discussed in \cite{conf:borgers2014event}. Event-triggered centralized state estimation for linear Gaussian systems is proposed in \cite{cite:shi2016linear}. Finally, the covariance intersection algorithm is investigated to enable a centralized event-triggered estimator \cite{conf:molin2015consistency}.

\subsection{Distributed Event-Triggered Estimation Algorithms}    

    As one of the main goals of sensor networks is to perform estimation in a distributed manner, event-triggered approaches are also applied in the corresponding distributed scenarios, including Kalman filters with covariance intersection \cite{conf:battistelli2016distributed}. An event-triggered distributed Kalman filter is proposed in \cite{conf:event_multisensor} for networked multi-sensor fusion systems with limited bandwidth. An event-triggered consensus Kalman filter with guaranteed stability is proposed in \cite{conf:eventguarantees}. The authors in \cite{conf:he-timebased,conf:he} propose an event-triggered communication protocol and provide a designing mechanism for triggering thresholds with guaranteed boundedness of the error covariance for distributed Kalman filters with covariance intersection. Notably, the send-on-delta data transmission mechanisms are proposed in the event-triggered Kalman consensus filters \cite{conf:li2016event}. Moreover, event triggering on the sensor-to-estimator and estimator-to-estimator channels are investigated with regard to the distributed Kalman consensus \cite{conf:zhang2017distributed}. Transmission delays and data drops in a distributed event-triggered control system are considered in \cite{conf:wang2011event}.
    Moreover, multiple distributed sensor nodes are considered in  \cite{conf:trimpe2015distributed}, where the sensors are used to observe a dynamic process and to exchange their measurements sporadically aiming to estimate the full state of the dynamic system. Significant deviation from the information predicted based on the last transmitted information is monitored to obtain a data-driven distributed Kalman filter \cite{conf:battistelli2016distributed}. For more details on the related work, we refer to the research presented on \cite{conf:zou2017event}.
    %Transmission delays and data drops in distributed event-triggered networked control system was considered in \cite{conf:wang2010asymptotic,conf:delay}. More analysis and proving the stability of the system given abound delays by given deadlines are presented in \cite{conf:wang2011event}. 

%\subsection{Event-Triggered Diffusion Kalman Filters}    

It is deemed not efficient to propose an event-triggered diffusion Kalman filter to reduce communication and computational costs and to require sharing the local covariance matrices with the purpose of finding the optimal diffusion weights. Therefore, we focus on event-triggered diffusion Kalman filters without sharing the local covariance matrices. This research direction has two related works. The first one is dedicated to a partial diffusion Kalman filter \cite{vahidpour2017partial}, which is mainly intended to address the diffusion step where Every wireless node shares only a subset of its intermediate estimate vectors among its neighbors at each iteration. However, there is no resource-saving achieved at the measurement update step, which already imposes high communication and sensing overheads. Also, it is not duty cycling the whole communication process within the diffusion step. On the other hand, the concern of the other work \cite{conf:chen2017diffusion} is optimizing the measurement update step while neglecting the diffusion step, which is associated with the significant overhead, as shown in \cite{vahidpour2017partial}. In the present study, we consider both the diffusion and the measurement steps; we temporarily shut-down the sensing and communication between nodes. To the best of our knowledge, the present paper is the first work focused on event-triggering the aforementioned steps of diffusion Kalman filter without the need to share the local covariance matrices. Moreover, we evaluate the proposed mechanism on a real testbed for localization and time synchronization.
%In addition, we do not depend on monitoring the change between the expected state and the calculated one.
    %We investigate the related work in the area.

%https://www3.nd.edu/~lemmon/projects/NSF-05-1518/

%\section{System Model} \label{sec:system}
\section{Triggering Logic Motivation}\label{sec:mot}
%Inherent in the original centralized Kalman filter a powerful tool, which is the error covariance matrix. 
One of the merits of the original centralized Kalman filter is the error covariance matrix. It is deemed as a perfect measure of the expected accuracy of the estimated state and can be utilized for regulating the resource consumption based on the particular application needs. However, when it comes to the distributed diffusion Kalman filter, we do not have local access to the error covariance matrix \cite{conf:diffusion-org}. Therefore, we aim to obtain the expected accuracy of the estimated state in the distributed diffusion Kalman filter, in which the local estimators do not have access to all measurements. To provide the step-by-step rationale behind the proposed approach, let us start by considering a background example to illustrate the relationship between the global and local covariance information at each node based on the partial access to the measurements, which is the case of distributed filtering. 

%Towards achieving our goal, let us start by a background example to state the relationship between the global covariance and the local covariance information at each node based on partial access to the measurements, which is the case for distributed filtering. 

%Let us motivate this goal by a background example.
%it is important to note that $P_{k,i|i}$ and $P_{k,i+1|i}$ do not represent the covariances of the state estimation errors $\Tilde{x}_{k,i|i}$ and $\Tilde{x}_{k,i+1|i}$ any longer, since the diffusion update in \eqref{equ:diffusion} combines the estimates from the neighbors and does not take into account the recursions of $P_{k,i|i}$ matrices. Thus, the covariance of the error is not available locally anymore, and we look for recovering the notion of the estimation error.
%\begin{equation}
%    \hat{x}_{k,i|i}= \sum\limits_{j\in\mathcal{N}_{k}} c^{k,j} \Psi_{j,i}. \label{equ:diffusion}
%\end{equation}
 %In this sub-section, we study how the local error covariance matrix to the global error covariance matrix aiming to recover the notion of error covariance. 
%, and it messes up the notion of the state error
\begin{example}
%As a background, 
Let us introduce $\hat{x}_1$ as the least-mean-squares estimator of $x$ given a zero-mean observation $y_1$, $\hat{x}_2$ as the least-mean-squares estimator of $x$ given a zero-mean observation $y_2$, and $\hat{x}$ as least-mean-squares estimator of $x$ given all observations. As a consequence, we obtain the two separate estimators for $x$ given two separate measurements and a global estimator given all the measurements. Let $P_1$, $P_2$ and $P$ denote the corresponding local and global error covariance matrices, respectively. We assume that the measurement noises are uncorrelated and have zero mean. It can be shown \cite[p.89]{conf:sayed2003fundamentals}
that the global and local error covariance matrices are related as per the following equation
\begin{equation}
    P^{-1} = P_1^{-1} + P_2^{-1} - R_x,
    \label{equ:p12inv}
\end{equation}
where $R_x$ is the positive-definite covariance matrix of $x$. 
\end{example}

%We are going to apply the same rule to the measurement update step. 
In the case of the distributed Kalman filter, every node obtains access to the measurements of its neighbors in addition to its local measurements. Therefore, we introduce two terms: \textit{individual} and \textit{local} estimates. The individual estimate considers only the measurements at the node itself without referring to its neighbors, while the local one considers the individual measurements together with those the measurements of its neighbors. More specifically, we denote the individual estimate by $\hat{x}^{\text{ind}}_{k,i|j}$, which corresponds to the optimal linear estimate of $x_i$ at time step $i$ given only the individual measurements up to time $j$ at node $k$ without considering its neighbors; and the individual error covariance matrix is denoted by $P^{\text{ind}}_{k,i|j}$. The local estimate at node $k$ is denoted by $\hat{x}^{\text{loc}}_{k,i|j}$ and corresponds to the optimal linear estimate of $x_i$ at time step $i$ given its individual measurements and measurements across the neighbors of node $k$ until time step $j$. The local error covariance matrix is denoted by $P^{\text{loc}}_{k,i|j}$. We also denote the global estimate by $\hat{x}_{i|j}$, which corresponds to the optimal linear estimate of $x_i$ given all observations across all nodes up to time step $j$ and its error covariance matrix by $P_{i|j}$. %Applying \eqref{equ:p12inv} on the measurement step results in:

The global error covariance matrix $P_{i|i}$ denotes the expected estimation error. Therefore, we can duty cycle collecting measurements and message exchange processes based on a threshold on the trace of the global error covariance matrix $P_{i|i}$. However, in the distributed Kalman filter algorithms, every node has access only to its local matrix $P^{\text{loc}}_{k,i|i}$ in distributed Kalman filter algorithms and does not have access to $P_{i|i}$ locally. Therefore, let us investigate if there is a direct relation between $P_{i|i}$ and $P^{\text{loc}}_{k,i|i}$. Instead of considering two nodes, we extend \eqref{equ:p12inv} to $N$ nodes where every node uses its individual measurements only \cite[p.14]{conf:diffusion} as follows:
\begin{equation}
    P_{i|i}^{-1} = \sum\limits_{k=1}^N (P^{\text{ind}}_{k,i|i})^{-1} - (N-1) \Pi^{-1}_i,
    \label{equ:ppind}
\end{equation}
where $\Pi_i$ is the covariance matrix of $x_i$. The individual error covariance matrices $P^{\text{ind}}_{k,i|i}$ are expected to decrease over time in observable systems. Therefore, their inverses in the first term $\sum\limits_{k=1}^N (P^{\text{ind}}_{k,i|i})^{-1}$ in \eqref{equ:ppind} become dominant. Therefore, $P_{i|i}^{-1}$ can be approximated as follows:

\begin{equation}
    P_{i|i}^{-1} \approx \sum\limits_{k=1}^N (P^{\text{ind}}_{k,i|i})^{-1}. \label{equ:ppindapprox}
\end{equation}
Then, let us apply \eqref{equ:p12inv} while sharing the measurements between the neighbors, i.e, considering the local estimates. We can find that the local error covariance $P^{\text{loc}}_{k,i|i}$ of each node $k$ depends on $P^{\text{ind}}_{l,i|i}$ of each neighbor $l$ as the local estimation process considers the neighbors measurements. Therefore, we can relate $P^{\text{loc}}_{k,i|i}$ to $P^{\text{ind}}_{k,i|i}$ in \eqref{equ:plocpind} with the aid of the adjacency matrix $A$ which has unity entry if the corresponding nodes are neighbors, and zero otherwise. Here $a_{l,k}$ denotes the element at row $l$ and column $k$ of matrix $A$ \cite[p.14]{conf:diffusion}.

\begin{equation}
    (P^{\text{loc}}_{k,i|i})^{-1} = \sum\limits_{l=1}^N a_{l,k} (P^{\text{ind}}_{l,i|i})^{-1} - \Big( \sum\limits_{l=1}^N a_{l,k} -1 \Big)\Pi^{-1}_i, \label{equ:plocpind}
\end{equation}
where
\begin{equation}
    a_{l,k} =
  \begin{cases}
    1       &  \text{if } l \in \mathcal{N}_{k}, \\
    0     &    \text{otherwise}. 
  \end{cases}
  \label{equ:Adj}
\end{equation}

If we sum up \eqref{equ:plocpind} for all $N$ nodes with a set of real weights $\gamma_k$ associated with every node $k$, we obtain the following combinations:

\begin{eqnarray}
   \sum\limits_{k=1}^N \gamma_k (P^{\text{loc}}_{k,i|i})^{-1} &=& \sum\limits_{l=1}^N \sum\limits_{k=1}^N \gamma_k a_{l,k} (P^{\text{ind}}_{l,i|i})^{-1} \nonumber\\
   &&- \Big( \sum\limits_{l=1}^N \sum\limits_{k=1}^N \gamma_k a_{l,k} -\sum\limits_{k=1}^N \gamma_k \Big)\Pi^{-1}_i. \label{equ:plocpindweights}
\end{eqnarray}

Setting the weights in \eqref{equ:plocpindweights} such that $\sum\limits_{k=1}^N \gamma_k a_{l,k}=1$ for all $l$, and having the first term again on the right-hand side dominant, as the individual error covariance matrices are expected to get smaller with time, results in

\begin{equation}
   \sum\limits_{k=1}^N \gamma_k (P^{\text{loc}}_{k,i|i})^{-1} \approx \sum\limits_{l=1}^N (P^{\text{ind}}_{l,i|i})^{-1}. \label{equ:plocpindapprox}
\end{equation}

Equating the two approximations in \eqref{equ:plocpindapprox} and \eqref{equ:ppindapprox} results in

\begin{equation}
   P_{i|i}^{-1} \approx \sum\limits_{k=1}^N \gamma_k (P^{\text{loc}}_{k,i|i})^{-1} .   \label{equ:pplocapprox}
\end{equation}

We have shown direct approximation between the matrix $P_{i|i}$ and local available matrix $P^{\text{loc}}_{k,i|i}$. Therefore, we can trigger collecting measurements based on the trace of the local error covariance matrix $P^{\text{loc}}_{k,i|i}$, which is available in distributed Kalman filters. This is the motivation behind the triggering logic considered in the present study.  

%%%%%%%%%%%%%%%%%%%%%%%%%%%%%%%%%%%%%%%%%%%%%%%%%%
 \begin{figure}[t]
\centering
\includegraphics[width = 0.45\textwidth]{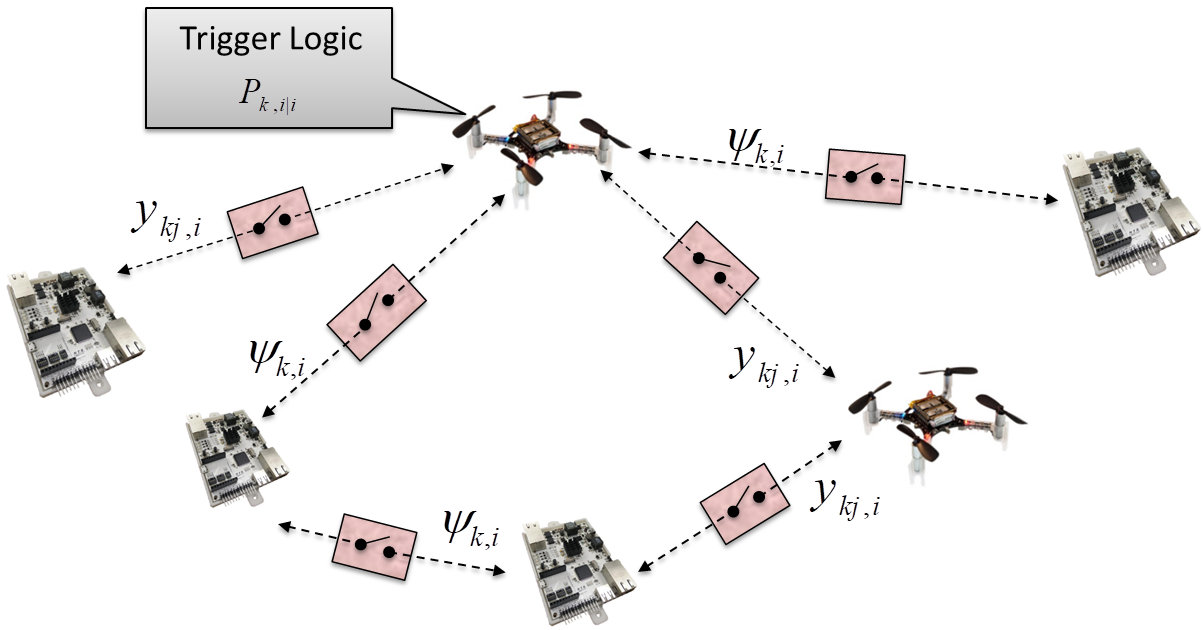}
\caption{
Every sensor node runs a distributed event-triggered state estimator to obtain the network state $x_{k,i|i}$. The nodes only share the measurements $y_{kj,i}$ and intermediate estimates $\Psi_{j,i}$ with their neighbors.} 
%The trigger logic is based on monitoring a local signal indicative of estimation error, thus linking the transmission and the sensing decisions to the estimation performance. 
%There is a triggering logic running on the leader node based on the network state of interest. The gray blocks constitute the event-based state estimator to be designed herein. The depicted idea can be applied in different scenarios where resources are limited.}
\label{fig:eventsys}
\vspace{-7mm}
\end{figure}

\section{Event-Triggered Diffusion Extended Kalman Filter Algorithm} \label{sec:algo}
We consider the event-triggered distributed state estimation problem over a network of $N$ distributed nodes indexed by $k\in \{0, \ldots, N-1\}$ as shown in Figure \ref{fig:eventsys}. Each node represents a sensor and an estimator. Moreover, we consider that two nodes are connected if they can communicate directly with each other. Let us consider the following nonlinear time-varying system to fit our nonlinear application as follows: 
\begin{equation}
\begin{split}
x_{i+1} &= f_i(x_{i}) + G_i n_{i}\\
y_{kj,i} &= h_{k,i}(x_{i}) + v_{k,i}\\
\end{split}
\label{eq:statespace}
\end{equation}
where $x_{i} \in \mathbb{R}^m$ is the state at time step $i$ and $y_{kj,i} \in \mathbb{R}^{L_k}$ is the measurement between node $k$ and its neighbor node $j$~$\in$~$\mathcal{N}_k$ at time step $i$. Furthermore, the process noise $n_i$ and the measurement noise $v_{k,i}$ are assumed to be uncorrelated, and zero mean white Gaussian noises. The matrices $Q_i$ and $R_i$ are the process and the measurement covariance matrices at time step $i$, respectively. The state update and measurement functions are denoted by $f_i$ and $h_{k,i}$, respectively. 

We denote the estimate at time step $i$ of $x_{i}$ by $\hat{x}_{k,i|s}$ given the observations up to time $s$ at node $k$, when every node seeks to minimize the mean squared error $\mathbb{E} \| x_{i}-\hat{x}_{k,i|i}\|^2$. To handle the nonlinearity in the considered model, we linearize \eqref{eq:statespace} at a linearization point $z$, and apply the diffusion Kalman filtering algorithm \cite{conf:diffusion-org,conf:diffusion}. The resulting state update and measurement functions are shown in \eqref{equ:linearize1} and \eqref{equ:linearize2}. The linearization clearly depends on $z$, and this point should be the best available local estimate of $x_i$: %Therefore, every node chooses $\psi=\hat{x}_{k,i|i}$ as the local linearization point. The complete linearization proof is in \cite{conf:nonlineardiff}. 
%\hat{H}_{kj,i}

\begin{align}
\bar{F}_i(z) &\delequal \frac{\partial f_i(x)}{\partial x}\Big\rvert_{x=z} \label{equ:linearize1},\\
\bar{H}_{k,i}(z) &\delequal \frac{\partial h_{k,i}(x)}{\partial x}\Big\rvert_{x=z}. \label{equ:linearize2}
%&\bar{u}_{k,i}(\psi) \delequal  f_i(\psi) - \bar{F}_i(\psi)\psi 
\end{align}

%We have the non-linearity solved by Equation \ref{eq:fh}. 

%Exposing diffusion distributed Kalman filter capabilities towards satisfying the system constraints using efficient duty cycling is our ultimate goal. Our proposed algorithm works well with the diffusion distributed Kalman filter. However, our application, namely, localization and time synchronization and many others require nonlinear time variant system model. Thus, we propose our algorithm to deal with the nonlinear time variant case which is the more general case as shown in Algorithm 1, where we have the non-linearity solved by Equation \ref{eq:fh}.
 
%We denote the conjugate transpose of $A$ by $A^T$. 

Given a linearized model, we subsequently explain the proposed event-triggered diffusion extended Kalman filter algorithm shown in Algorithm $1$. One of the nodes is elected beforehand as a leader based on the accessibility of important measurements, which facilitates reaching the best local estimate compared to the followers. We denote the leader node by subscript $L$ and its job to take the duty-cycling decision. Choosing a proper leader is crucial in saving energy; however, the election process based on the available estimates is out of scope of this paper. Algorithm $1$ is started with the measurement update (step 1), in which every node $k$ obtains a local estimate $\Psi_{k,i}$ at time step $i$. Then, information from the neighbors of node $k$ is diffused in a convex combination to produce a better new state estimate in step $2$. The $c_{k,j}$ elements represent the weights that are used by the diffusion algorithm to combine neighboring estimates. Step 1 and 2 are only executed if the trace of the required part of the leader matrix $P_{L,i|i-1}$ is greater than the user-defined threshold $\pi_{\max}$. Explicitly, the triggering event is defined as $\tr(WP_{L,i|i-1}W^{\T}) > \pi_{\max}$, where $W$ is a weighting matrix to choose the required part of $P_{L,i|i-1}$. If the triggering event is not satisfied, we do not take measurements $y_{kj,i}$, and we save the communication costs at steps $1$ and $2$. Instead, we perform the propagation update (step $3$), in which every node considers the new estimates as the old available ones $\hat{x}_{k,i|i} = \hat{x}_{k,i|i-1}$ and its corresponding local matrix $P_{k,i|i}=P_{k,i|i-1}$. Finally, every node performs the time update (step 4) in all cases. 

It should be noted that the analysis presented in Section \ref{sec:mot} shows the relationship between the global error covariance and the available local error covariance during the measurement update (step 1) in Algorithm $1$. However, the diffusion update (step 2) does not take into account recursions for these local error covariance matrices as it only combines the estimates of the neighbors without considering their local error covariance matrices. Moreover, exchanging the $P^{\text{loc}}_{k,i|i}$ between neighbors to maintain the exact expected estimation error is of a great overhead in sensor networks. Furthermore, the diffusion step decreases the estimation error so that it should be included. Therefore, we continue with the modified version of $P^{\text{loc}}_{k,i|i}$, which we call \textit{diffusion error covariance} matrix, and denote it by $P_{k,i|i}$ in Algorithm $1$.

%The time synchronization and localization problem estimates the clock parameters and the 3D position using the timestamp measurements and the model of the system. 
%Our algorithm aims to enhance the DKAL algorithm in \cite{conf:d-slats} by reducing the computation and the communication overheads to fit low power embedded devices. 
%repeated
% $Q_t$ and $R_{j,t}$ are the process covariance matrix at time $t$, and the measurement noise covariance matrix of node $j$ at time $t$, respectively. 

%this is moved to evaluation
%While we focus on the localization accuracy in this scenario, we use $W$ as 
%   \[
%      W = \begin{bmatrix}
%         1 & 0 & 0 & 0 & 0\\
%         0 & 1 & 0 & 0 & 0\\
%         0 & 0 & 1 & 0 & 0
%      \end{bmatrix}.
%   \]

\begin{algorithm}[ht]
\begin{flushleft}
%\begin{algorithmic}
\caption{Event-Triggered Diffusion Extended Kalman Filter}
\label{alg:one}
Start with $\hat{x}_{k,0|-1}=x_0$ and $P_{k,0|-1}= \Pi_0$ for all $k$, and at every time instant $i$, compute at every node $k$:\\
\If{$\tr(WP_{L,i|i-1}W^T) > \pi_{\max}$}
{
\textbf{Step 1}: Measurement update:
\begin{eqnarray}
\hat{H}_{kj,i}&=& \bar{H}_{j,i}(\hat{x}_{k,i|i-1})\label{eqn:1}\\
%P_{k,t|t}^{-1} &=& \tilde{Q}_N \nonumber\\
P^{-1}_{k,i|i}&=& P^{-1}_{k,i|i-1} + \sum\limits_{j\in\mathcal{N}_{k}}\hat{H}_{kj,i}^T {R}^{-1}_i \hat{H}_{kj,i} \label{eqn:p_nonlinear}\\ %\label{eqn:2}
\Psi_{k,i}&=& \hat{x}_{k,i|i-1}  \nonumber \\
 &&+ P_{k,i|i}\sum\limits_{j\in\mathcal{N}_{k}}\hat{H}^T_{kj,i}R_{i}^{-1}[y_{kj,i}-h_{j,i}(\hat{x}_{k,i|i-1})]  \nonumber \label{eqn:psi_nonlinear} \\ 
\end{eqnarray} \nonumber
\textbf{Step 2}: Diffusion update: %\nonumber\\
\begin{eqnarray}
\hat{x}_{k,i|i}&= &\sum\limits_{j\in\mathcal{N}_{k}} c_{k,j} \Psi_{j,i} \label{equ:diff} 
\end{eqnarray}
}
\Else{
\textbf{Step 3}: Propagation update:
\begin{eqnarray}
\hat{x}_{k,i|i} &=& \hat{x}_{k,i|i-1} \label{equ:x_propupdate} \\
P_{k,i|i} &=& P_{k,i|i-1}  
\end{eqnarray}
}
\textbf{Step 4}: Time update:
\begin{eqnarray}
%\hat{x}_{k,i+1|i} &= &\bar{F}_{i}(\hat{x}_{k,i|i}) \hat{x}_{k,i|i} + \bar{u}_{k,i}(\hat{x}_{k,i|i})\nonumber\\
\hat{x}_{k,i+1|i} &= & f_i(\hat{x}_{k,i|i}) \\
P_{k,i+1|i} &= &\bar{F}_{i}(\hat{x}_{k,i|i})P_{k,i|i}\bar{F}_{i}^T(\hat{x}_{k,i|i}) + G_i Q_{i} G_i^T 
%P_{k,i+1|i} &= &\bar{F}_{i}(\hat{x}_{k,i|i})P_{k,i|i}\bar{F}_{i}(\hat{x}_{k,i|i})^{*} + G_{i}Q_{i}G_{i}^{*}
\end{eqnarray}
%\end{algorithmic}
\vspace{-7mm}
\end{flushleft}
\end{algorithm}

   % resource concern

\section{Theoretical Analysis} \label{sec:analysis}

%We start by giving the motivation behind our triggering logic in Algorithm 
%We show that our algorithm is unbiased estimator. Finally, we drive the relationship between $P_{k,i|i}$ and the augmented error covariance. 

%Although the extended Kalman filter (EKF) has proven to work appropriately in many nonlinear practical applications, its general convergence guarantees cannot be proved even in the centralized version  \cite{conf:ekf-convergence}. Thus, 

We limit our analysis to the linear case. We prove that event-triggered diffusion Kalman is an unbiased estimator. Next, we show the relationship between the local matrix $P_{k,i|i}$ and the augmented error covariance. Let us consider the following linear time-varying system 
\begin{align}
x_{i+1} &= F_ix_{i} + G_i n_{i},\label{equ:xlinear}\\
y_{kj,i} &= H_{k,i} x_{i} + v_{k,i}.\label{equ:ylinear}
%\label{eq:statespacelinear}
\end{align}

\begin{lemma}
The event-triggered diffusion Kalman filter is an unbiased estimator.
\end{lemma}
\begin{proof}
The measurement update step in the linear case results in the following (linear version of \eqref{eqn:p_nonlinear}, \eqref{eqn:psi_nonlinear}):
\begin{align}
P^{-1}_{k,i|i}& = P^{-1}_{k,i|i-1} + \sum\limits_{j\in\mathcal{N}_{k}} H_{j,i}^T {R}^{-1}_i H_{j,i},
\label{equ:pinv} \\
\Psi_{k,i} & = \hat{x}_{k,i|i-1} +  P_{k,i|i}\sum\limits_{j\in\mathcal{N}_{k}} H^T_{j,i}R_{i}^{-1}[y_{kj,i}-H_{j,i}\hat{x}_{k,i|i-1}].\label{equ:psi}
\end{align}
\vspace{-5mm}
The estimation error $\tilde{x}_{k,i|i-1}$ at the end of Algorithm $1$ is defined and updated according to the formula:
\begin{align}
\tilde{x}_{k,i|i-1} &\delequal x_i-\hat{x}_{k,i|i-1} \nonumber\\
&= F_{i-1} \tilde{x}_{k,i-1|i-1} + G_{i-1}n_{i-1}. \label{equ:xerrupdate}
\end{align}

After defining the estimation error at the end of the measurement update by $\tilde{\Psi}_{k,i}$, we obtain that:
\begin{eqnarray}
 \tilde{\Psi}_{k,i} &\delequal&  x_i- \Psi_{k,i} \nonumber\\
&\stackrel{\eqref{equ:psi}}{=}& \tilde{x}_{k,i|i-1} - P_{k,i|i}\sum\limits_{j\in\mathcal{N}_{k}}H^T_{j,i}R_{i}^{-1}\Big[y_{kj,i}-H_{j,i}( x_i - \tilde{x}_{k,i|i-1}) \Big] \nonumber \\
&\stackrel{\eqref{equ:ylinear}}{=}& \tilde{x}_{k,i|i-1} - P_{k,i|i}\sum\limits_{j\in\mathcal{N}_{k}} H^T_{j,i}R_{i}^{-1} \Big[ H_{j,i}\tilde{x}_{k,i|i-1} + v_{j,i} \Big] \nonumber \\
&=& P_{k,i|i}\Big(P_{k,i|i}^{-1}-\sum\limits_{j\in\mathcal{N}_{k}} H^T_{j,i}R_{i}^{-1}H_{j,i}\Big)\tilde{x}_{k,i|i-1}  \nonumber \\
&& -  P_{k,i|i}\sum\limits_{j\in\mathcal{N}_{k}} H^T_{j,i}R_{i}^{-1}  v_{j,i}.
\label{equ:psierr1}
\end{eqnarray}

Using \eqref{equ:pinv} to simplify \eqref{equ:psierr1} results in the following: 
\begin{equation}
  \tilde{\Psi}_{k,i} \stackrel{\eqref{equ:pinv}}{=} P_{k,i|i}P_{k,i|i-1}^{-1}\tilde{x}_{k,i|i-1} - P_{k,i|i}\sum\limits_{j\in\mathcal{N}_{k}} H^T_{j,i}R_{i}^{-1}  v_{j,i}. \label{equ:psierr2}
\end{equation}

Applying the diffusion step \eqref{equ:diff} results in 
\begin{align}
      \tilde{x}_{\s k,i|i} &= \sum_{l\in\mathcal{N}_k} c_{l,k}\tilde{\Psi}_{l,i} \nonumber\\
          &\stackrel{\eqref{equ:psierr2}}{=} \sum_{l\in\mathcal{N}_k} c_{l,k} \bigg[ P_{\s l,i|i}P_{\s l,i|i-1}^{-1}\tilde{x}_{\s l,i|i-1} - P_{\s l,i|i}\sum\limits_{j\in\mathcal{N}_{l}} H^T_{j,i}R_{i}^{-1}  v_{j,i}  \bigg].\label{equ:differr}
\end{align}
   
Executing $M$ time updates and propagation updates before $\tr(WP_{L,i|i}W^T)$ exceeds the threshold $\pi_{\max}$ results in the following:
\begin{eqnarray}
\tilde{x}_{\s k,i+M+1|i+M} &\stackrel{\eqref{equ:xerrupdate}}{=}& F_{\s i+M} \tilde{x}_{\s k,i+M|i+M} + G_{\s i+M} n_{\s i+M} \nonumber\\
 &\stackrel{\eqref{equ:x_propupdate}}{=}& F_{\s i+M} \tilde{x}_{\s k,i+M|i+M-1} + G_{\s i+M} n_{\s i+M}\nonumber\\
&\stackrel{\eqref{equ:xerrupdate}}{=}& F_{\s i+M} \Big( F_{\s i+M-1} \tilde{x}_{\s k,i+M-1|i+M-1}\nonumber\\
&&+ G_{\s i+M-1} n_{\s i+M-1} \Big) + G_{\s i+M} n_{\s i+M}.\nonumber
\end{eqnarray}

After using \eqref{equ:xerrupdate} and \eqref{equ:x_propupdate} multiple times to obtain $\tilde{x}_{\s k,i+M+1|i+M}$ as a function of $\tilde{x}_{\s k,i|i}$, we obtain that:
\begin{eqnarray}
\tilde{x}_{\s k,i+M+1|i+M} &=& \prod_{j=0}^{M} F_{\s i+M-j} \tilde{x}_{\s k,i|i} + \sum_{l=1}^{M} \prod_{j=0}^{M-l} F_{\s i+M-j} G_{\s i+l-1} n_{i+l-1} \nonumber\\
&&+ G_{\s i+M} n_{\s i+M}. \label{equ:timeequalerr}
\end{eqnarray}
%\stackrel{\eqref{equ:xerrupdate}}{=} \hdots \stackrel{\eqref{equ:x_propupdate}}{=} \hdots \nonumber\\
%&\stackrel{\eqref{equ:xerrupdate}}{=}

Inserting \eqref{equ:differr} in \eqref{equ:timeequalerr} results in 
\begin{align}
\tilde{x}_{\s k,i+M+1|i+M} =& \prod_{j=0}^{M} F_{\s i+M-j} \Bigg[\sum_{l\in\mathcal{N}_k} c_{\s l,k} \bigg[ P_{\s l,i|i}P_{\s l,i|i-1}^{-1}\tilde{x}_{\s l,i|i-1} \nonumber\\
& - P_{\s l,i|i}\sum\limits_{j\in\mathcal{N}_{l}} H^T_{\s j,i}R_{\s i}^{-1}  v_{\s j,i}  \bigg]\Bigg] \nonumber\\
& + \sum_{l=1}^{M} \prod_{j=0}^{\s M-l} F_{\s i+M-j} G_{\s i+l-1} n_{\s i+l-1} + G_{\s i+M} n_{\s i+M}. \label{equ:allerr}
\end{align}

Taking the expectations of both sides of \eqref{equ:allerr} leads to in the following recursion given that we have zero mean noises:
\begin{equation}
\mathbb{E} \tilde{x}_{\s k,i+M+1|i+M} = \prod_{j=0}^{M} F_{\s i+M-j} \sum_{l\in\mathcal{N}_k} c_{\s l,k}  P_{\s l,i|i}P_{\s l,i|i-1}^{-1}\mathbb{E} \tilde{x}_{\s l,i|i-1}.
\end{equation}

Since $\mathbb{E} \tilde{x}_{l,0|-1}=0$ as $\hat{x}_{l,0|-1}=0$ and $\mathbb{E}x_0=0$ \cite{conf:diffusion}, we conclude that the event-triggered diffusion Kalman is an unbiased estimator.
\end{proof}

 The diffusion step \eqref{equ:diff} in Algorithm 1 combines the intermediate estimates from neighbors without combining the corresponding error covariance matrices, as we mentioned before. Therefore, we need to identify the new relationship between $P_{k,i|i}$, and the augmented error covariance. We define the augmented state-error vector $\tilde{\mathcal{X}}_{i|i}$ for the whole network as follows:
\begin{equation}
      \tilde{\mathcal{X}}_{\s i|i} \delequal \begin{bmatrix}
         \tilde{x}_{\s 1,i|i}, \hdots , \tilde{x}_{\s N,i|i}
      \end{bmatrix}^T. \label{equ:xaug}
\end{equation}

The Kronecker product and vector of ones are denoted by $\otimes$ and $\mathbbm{1}$, respectively. The identity matrix of the size $N \times N$ is denoted by $I_N$. We further introduce the following block-diagonal matrices and $v_i$:
   \begin{align*}
      \mathcal{H}_i &\delequal \diag(H_{\s 1,i},\dots, H_{\s N,i}), \\
      \mathcal{P}_{\s i|i} &\delequal \diag(P_{\s 1,i|i}, \dots, P_{\s N,i|i}), \\
      \mathcal{P}_{\s i|i-1} &\delequal \diag(P_{\s 1,i|i-1}, \dots, P_{\s N,i|i-1}),\\
      v_{\s i} & \delequal [v_{\s 1,i}, \hdots, v_{\s N,i}]^T.
   \end{align*}

\begin{lemma}   
The relationship between the error covariance $\mathcal{P}_{\s \tilde{\mathcal{X}}|i} =\mathbb{E}(\tilde{\mathcal{X}}_{\s i|i}^{} \tilde{\mathcal{X}}_{\s i|i}^T)$ of the augmented state and the diffusion error covariance $P_{k,i|i}$ is defined as follows:
\begin{align}
  \mathcal{P}_{\s \tilde{\mathcal{X}}|i+M+1} &=  A_{\s i} \mathcal{P}_{\s \tilde{ \mathcal{X}}|i} A_{\s i}^T +  B_i \big( \mathbbm{1}\mathbbm{1}^T \otimes Q_{\s i+M}\big) B_{\s i}^T \nonumber\\
 &+\sum_{l=1}^{M} D_{\s i,l} \big(\mathbbm{1} \mathbbm{1}^T \otimes Q_{\s i+l-1} \big) D_{\s i,l}^T +E_{\s i}  R_{\s i+M+1} E_{\s i}^T, \label{equ:pfinal}
\end{align}
where
\begin{align}
    Z_{\s i} &\delequal \mathcal{C}^T \mathcal{P}_{\s i+M+1|i+M+1}  \mathcal{P}_{\s i+M+1|i+M}^{-1}, \nonumber\\
    A_{\s i} &\delequal Z_{\s i} \big( I_{\s N} \otimes \prod_{j=0}^{M} F_{\s i+M-j}\big), \nonumber\\ 
    B_{\s i} &\delequal Z_{\s i} \big(I_{\s N} \otimes G_{\s i+M} \big), \nonumber\\
    D_{i,l} &\delequal Z_{\s i}  \big( I_{\s N} \otimes \prod_{j=0}^{\s M-l} F_{\s i+M-j} \big) \big( I_{\s N} \otimes G_{\s i+l-1}\big),  \nonumber\\
    E_{\s i} &\delequal \mathcal{C}^T \mathcal{P}_{\s i+M+1|i+M+1} \mathcal{A}^T \mathcal{H}^T_{\s i+M+1} \mathcal{R}_{\s i+M+1}^{-1}.
\end{align}
\end{lemma}   
\begin{proof}
   %Following \cite[p.4]{conf:diffusion-org} in getting the augmented state-error update results in 
   Extending \eqref{equ:differr} to the augmented version results in
  \hspace{-5mm}   \begin{align*}
      \tilde{\mathcal{X}}_{\s i+M+1|i+M+1} \hspace{-2mm} \stackrel{\eqref{equ:xaug}}{=}&   \begin{bmatrix}
         \tilde{x}_{\s 1,i+M+1|i+M+1}, \hdots , \tilde{x}_{\s N,i+M+1|i+M+1}
      \end{bmatrix}^T \\
      \stackrel{\eqref{equ:differr}}{=}& \mathcal{C}^T
     \begin{bmatrix}
         P_{\s 1,i+M+1|i+M+1}P_{\s 1,i+M+1|i+M}^{-1}\tilde{x}_{\s 1,i+M+1|i+M} \\
         \vdots \\
         P_{\s N,i+M+1|i+M+1}P_{\s N,i+M+1|i+M}^{-1} \tilde{x}_{\s N,i+M+1|i+M}
      \end{bmatrix} \\
    & -\mathcal{C}^T \mathcal{P}_{\s i|i}  \mathcal{A}^T
      \begin{bmatrix}
         H_{\s 1,i+M+1}R_{\s 1,i+M+1}^{-1} v_{\s 1,i+M+1} \\
         \vdots \\
         H_{\s N,i+M+1}R_{\s 1,i+M+1}^{-1} v_{\s N,i+M+1} 
      \end{bmatrix}, 
   \end{align*}
   or equivalently
   \begin{align}
    \tilde{\mathcal{X}}_{\s i+M+1|i+M+1} =& {\ } \mathcal{C}^T \mathcal{P}_{\s i+M+1|i+M+1} \Big( \mathcal{P}_{\s i+M+1|i+M}^{-1}  \tilde{\mathcal{X}}_{\s i+M+1|i+M} \nonumber\\
    & - \mathcal{A}^T \mathcal{H}^T_{\s i+M+1} \mathcal{R}_{\s i+M+1}^{-1} v_{\s i+M+1}\Big). \label{equ:xim1}
   \end{align}
   with 
   \begin{equation}
\mathcal{C} \delequal C \otimes I_m \quad \quad \mathcal{A} \delequal A \otimes I_m,
  \end{equation}
   %$I_m$ is the identity matrix with size $m \times m$.
      where the element at row $l$ and column $k$ of diffusion matrix $C$  is $c_{\s l,k}$ in \eqref{equ:diff}. The adjacency matrix is defined as $A$ in \eqref{equ:Adj}. The size of $x_i$ in \eqref{eq:statespace} is $m$.  Similarly, extending \eqref{equ:timeequalerr} to the augmented version results in the following:
\begin{align}
  \tilde{\mathcal{X}}_{\s i+M+1|i+M} & = \big( I_{\s N} \otimes \prod_{j=0}^{M} F_{\s i+M-j}\big) \tilde{\mathcal{X}}_{i|i} 
  + \big(I_{\s N} \otimes G_{\s i+M} \big) \big( \mathbbm{1} \otimes n_{\s i+M}\big) \nonumber\\
 &+ \sum_{l=1}^{M} \big( I_{\s N} \otimes \prod_{j=0}^{M-l} F_{\s i+M-j} \big) \big( I_{\s N} \otimes G_{\s i+l-1}\big) \big(\mathbbm{1} \otimes n_{\s i+l-1} \big). 
 \label{equ:xim2}
\end{align}

Inserting \eqref{equ:xim2} into \eqref{equ:xim1} leads to the following:
\begin{eqnarray}
 \tilde{\mathcal{X}}_{\s i+M+1|i+M+1} &=& A_{\s i} \tilde{\mathcal{X}}_{\s i|i} +  B_{\s i} \big( \mathbbm{1} \otimes n_{\s i+M}\big) \nonumber\\
 &&+\sum_{l=1}^{M} D_{\s i,l} \big(\mathbbm{1} \otimes n_{\s i+l-1} \big) -E_{\s i}  v_{\s i+M+1}. \label{equ:ximfinal}
\end{eqnarray}
%Let $\mathcal{P}_{\tilde{\mathcal{X}}|i} =\mathbb{E}( \tilde{\mathcal{X}}_{i|i}\tilde{\mathcal{X}}^T_{i|i})$ denotes the covariance of the augmented state error $\tilde{\mathcal{X}}_{i|i}$. 
Taking the expectation of both sides of \eqref{equ:ximfinal} results in \eqref{equ:prepfinal} with the assumption that the state error $\tilde{\mathcal{X}}$, the time instances of modeling noise $n_i$, and the time instances of measurements noise $v_i$ are mutually independent \cite{conf:diffusion}. 

\begin{align}
  \mathcal{P}_{\tilde{\s \mathcal{X}}|i+M+1} =& {\ } \mathbb{E}\Big(\tilde{\mathcal{X}}_{\s i+M+1|i+M+1}^{} \tilde{\mathcal{X}}_{\s i+M+1|i+M+1}^T\Big)\nonumber\\
  \stackrel{\eqref{equ:ximfinal}}{=}& A_{\s i} \mathbb{E}\Big(\tilde{\mathcal{X}}_{\s i|i} \tilde{\mathcal{X}}_{\s i|i}^T\Big)A_{\s i}^T +  B_{\s i} \mathbb{E} \Big(\big( \mathbbm{1} \otimes n_{\s i+M}\big)\big( \mathbbm{1} \otimes n_{\s i+M}\big)^T \Big) B_{\s i}^T \nonumber\\
 &+ \sum_{l=1}^{M} D_{\s i,l} \mathbb{E}\Big(\big(\mathbbm{1} \otimes n_{\s i+l-1} \big)\big(\mathbbm{1} \otimes n_{\s i+l-1} \big)^T\Big)D_{\s i,l}^T \nonumber\\
 &+ E_{\s i}  \mathbb{E}\Big(v_{\s i+M+1}v_{\s i+M+1}^T\Big) E_{\s i}^T.  \label{equ:prepfinal}
 \end{align}
 Applying the property of Kronecker products that $(A \otimes B)(C \otimes D)^T=(A C^T \otimes B D^T)$ results in
  \begin{align}
  \mathcal{P}_{\tilde{\s \mathcal{X}}|i+M+1} 
  =& A_i \mathcal{P}_{\tilde{\s \mathcal{X}}|i} A_i^T +  B_i \big( \mathbbm{1}\mathbbm{1}^T \otimes Q_{\s i+M}\big) B_{\s i}^T \nonumber\\
 &+\sum_{l=1}^{M} D_{\s i,l} \big(\mathbbm{1} \mathbbm{1}^T \otimes Q_{\s i+l-1} \big) D_{\s i,l}^T +E_i  R_{\s i+M+1} E_{\s i}^T.  \label{equ:pfinal}
\end{align}
%relation \eqref{equ:pfinal} 
The above formula connects the error covariance $\mathcal{P}_{\tilde{\mathcal{X}}|i}$ of the augmented state and the diffusion error covariance matrix $P_{k,i|i}$ embedded in $\mathcal{P}_{\s i|i}$ and this concludes the proof.
\end{proof}
\section{Evaluation} \label{sec:eval}

%We are considering mainly the communication overhead and its associated accuracy with different network topologies to show the effectiveness of our proposed algorithm. 
In this section, we initiate with describing the application of the proposed approach to the considered localization and time synchronization problem. Thereafter, we present the experimental setup, in which we conducted the experiments. Finally, we perform case studies to obtain equitable evaluation results.

\subsection{Application to Localization and Time Synchronization}\label{sec:app}

 %conf:d-slats
One of the illustrative applications of the event-triggering body of work is the distributed localization and time synchronization problem, due to its excessive communication and computational overhead. 
%Moreover, these types of algorithms are quite useful among sensors network device where the finite resources is a great concern. 
Therefore, we consider this application to demonstrate the practicality of the proposed algorithm. The state vector consists of the three-dimensional position vector $\bm{p}_{k,i}$, the clock time offset $o_{k,i}$, and the clock frequency bias $b_{k,i}$ for all nodes. We adopt convention in which both $o_{k,i}$ and $b_{k,i}$ are described with respect to the global time clock, which is usually the clock of the leader node. Every node is interested in obtaining the state of the whole network. Therefore, the state vector is $x_{\s k,i}=[\bar{x}_{\s 1,i},...,\bar{x}_{\s N,i}]^T$, where
 $\bar{x}_{\s k,i} =\left[\bm{p}_{\s k,i}^T,\: o_{\s k,i},\: b_{\s k,i}\right]^T$.

%\begin{eqnarray}
%\bar{x}_{\s k,i} &= &\begin{bmatrix}
%\bm{p}_{\s k,i}^T\\
%o_{\s k,i}\\
%b_{\s k,i}
%\end{bmatrix}.
%\end{eqnarray}

The clock parameters evolve according to the first-order affine approximation of the dynamics $o_{\s k,i+1} = o_{\s k,i} + b_{\s k,i} \delta_i$ and $b_{\s k,i+1} = b_{\s k,i}$, where $\delta_i \delequal t_{\s L,i+1}-t_{\s L,i}$ given that $t_{\s L,i}$ is the time according to the leader node, which is the global time. Therefore, we can define the update function as follows:

\begin{eqnarray}
f_i(\bar{x}_{\s k,i}) &= &\begin{bmatrix}
\bm{p}_{\s k,i}\\
o_{\s k,i}+b_{\s k,i}\delta_{\s i}\\
b_{\s k,i}
\end{bmatrix}.
\end{eqnarray}

The proposed framework supports the three types of measurements which are distinguished by the number of messages exchanged between a pair of nodes. The measurement vector sent from node $j \in \mathcal{N}_k$ to node $k$ has the form $y_{\s kj,i}=\left[ d_{\s kj,i}, r_{\s kj,i}, \Gamma_{\s kj,i}\right]^T$,
%\begin{eqnarray}
%y_{\s kj,i} &= &\begin{bmatrix}
%d_{\s kj,i}\\
%r_{\s kj,i}\\
%\Gamma_{\s kj,i}
%\end{bmatrix},
%\end{eqnarray}
where, $d_{\s kj,i}$, represents the counter difference at time step $i$ which is the difference between the clock offsets of the two nodes $k$ and $j$. In turn, $r_{\s kj,i}$ represents a noisy measurement due to frequency bias discrepancies between $k$ and $j$, which is formally represented by single-sided two-way range. Finally, $\Gamma_{\s kj,i}$ is another distance measurement between nodes $k$ and $j$ based on a trio of messages between the nodes at time index $i$. This is a more accurate estimate than $r_{\s kj,i}$ owing to the mitigation of frequency bias errors from the additional message. It is formally called double-sided two-way range \cite{conf:d-slats}. We note that a subset of these measurements may be used rather than the full set, i.e., we can conduct experiments involving only $r_{\s kj,i}$, $\Gamma_{\s kj,i}$, or $d_{\s kj,i}$. The response time duration between the first pair of timestamps is denoted by $T_{\s RSP1}$. The measurement function is defined as follows \cite{conf:d-slats}:

%Time stamps $t_0(i)$ through $t_5(i)$ denote the locally measured transmission (TX) and reception (RX) times stamps, and 
%$T_{RSP}(i)$ and $T_{RND}(i)$ define, respectively, the response and the round-trip durations between the appropriate pair of these timestamps. The propagation velocity of radio is taken to be the speed of light in a vacuum, denoted by $c$. 

\begin{align}
 h_{\s k,i}(\bar{x}_{\s j,i})&=\left[\begin{array}{c}
\left(o_{\s j,i}-o_{\s k,i}\right) + \frac{1}{c}\left\Vert \bm{p}_{\s j,i}-\bm{p}_{\s k,i}\right\Vert _{\s 2}\\
\left\Vert \bm{p}_{\s j,i}-\bm{p}_{\s k,i}\right\Vert _{2}+\frac{c}{2}\left(b_{\s j,i}-b_{\s k,i}\right)T_{\s RSP1}\\
\left\Vert \bm{p}_{\s j,i}-\bm{p}_{\s k,i}\right\Vert _{\s 2} + c \tilde{\Gamma}_{\s kj,i}\\
\end{array}\right]
\label{eq:measmodel}
\end{align}

%where
%{\small \begin{align*}
%&\tilde{\Gamma}^{k,j}_i := \\ &\dfrac{(b^k_i-b^j_i)(T_{RND0}(i)T_{RND1}(i) - T_{RSP0}(i)T_{RSP1}(i))}{ (1+ b^k_i-b^j_i)T_{RND0}(i) + T_{RND1}(i) + T_{RSP0}(i) + ( 1 + b^k_i-b^j_i)T_{RPS1}(i)}
%\end{align*}}

\begin{figure*}[t]
\centering
\includegraphics*[width = 0.9\textwidth]{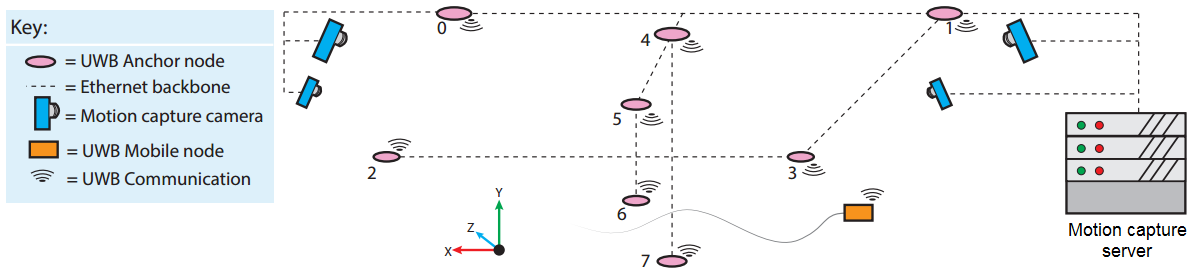}
\caption{Experimental setup, including, UWB anchor nodes, motion capture cameras, and UWB quadrotor nodes.}
\label{fig:full_sys}
\vspace{-4mm}
\end{figure*}

\begin{figure*}[t!]
\centering
\begin{minipage}[t]{0.30\textwidth}%
\includegraphics[height=1.56in]{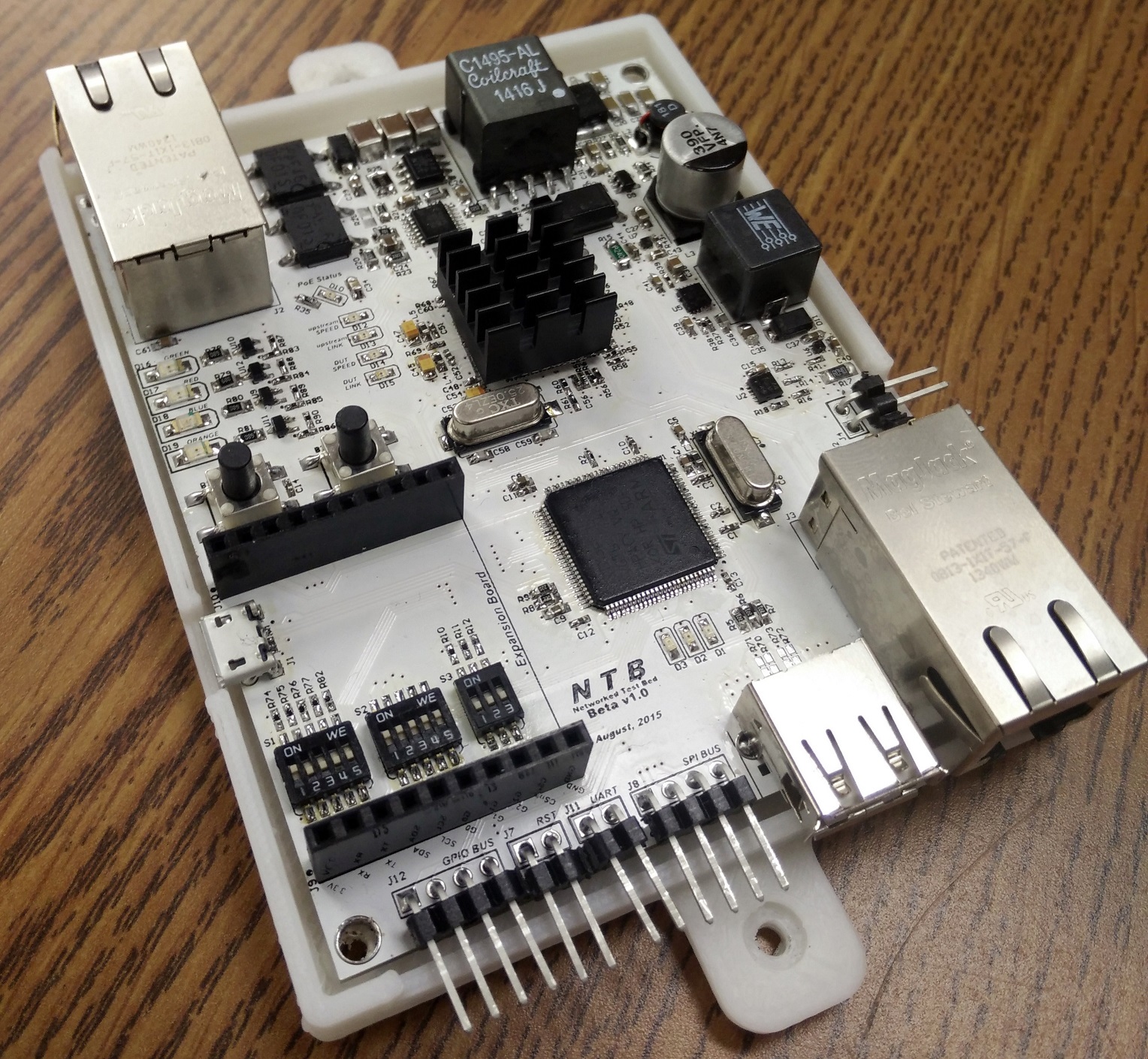}
\caption{Custom anchor with ARM Cortex M4 processor and UWB slot.}
\label{fig:ntbanon}
\end{minipage}\hspace{4mm}
\begin{minipage}[t]{0.30\textwidth}%
\centering
\includegraphics[height=1.56in]{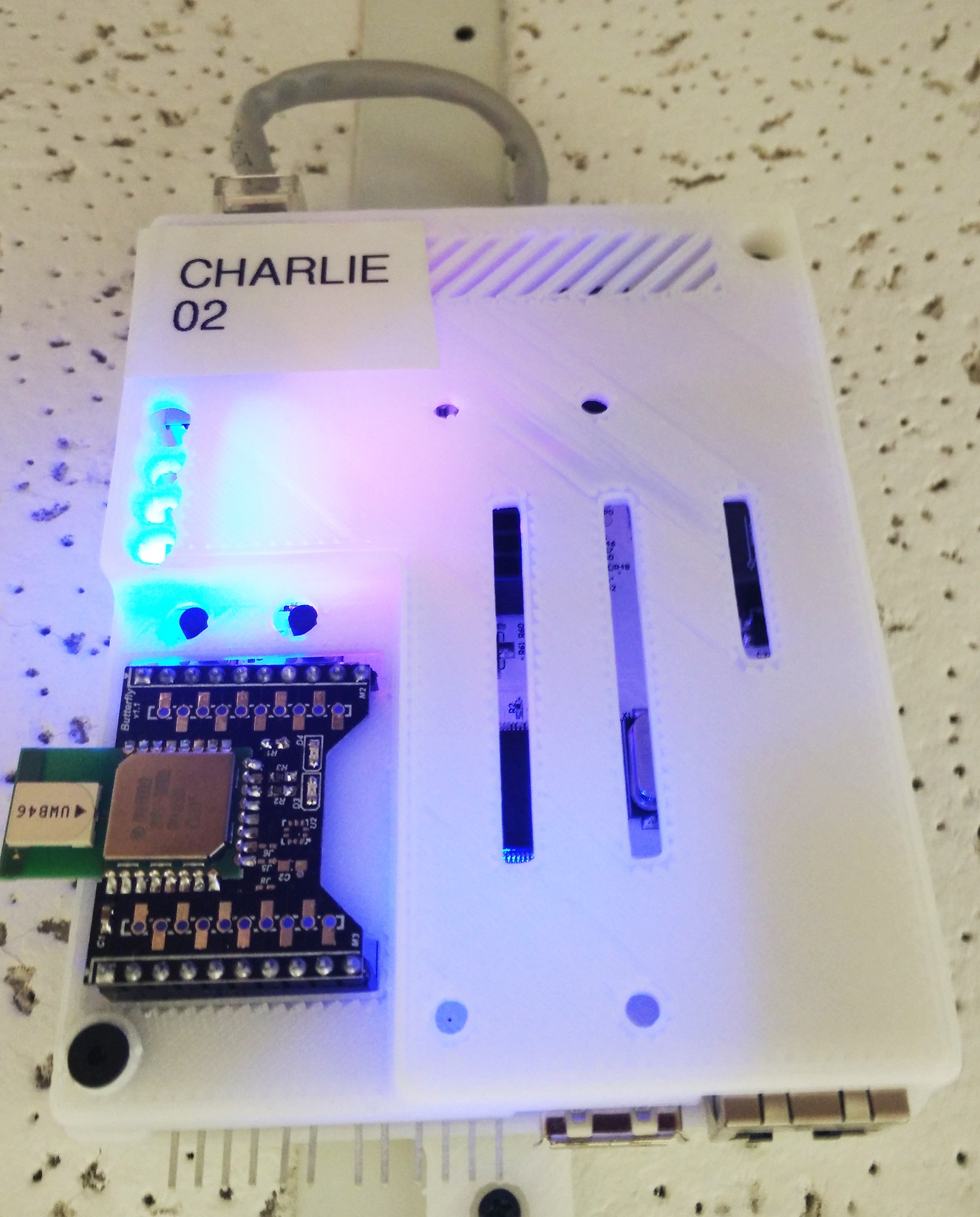}
\caption{Ceiling-mounted anchor with UWB radio in 3D-printed enclosure.}
\label{fig:ntbceil}
\end{minipage}\hspace{4mm}
\begin{minipage}[t]{0.30\textwidth}%
\centering
\includegraphics[height=1.56in]{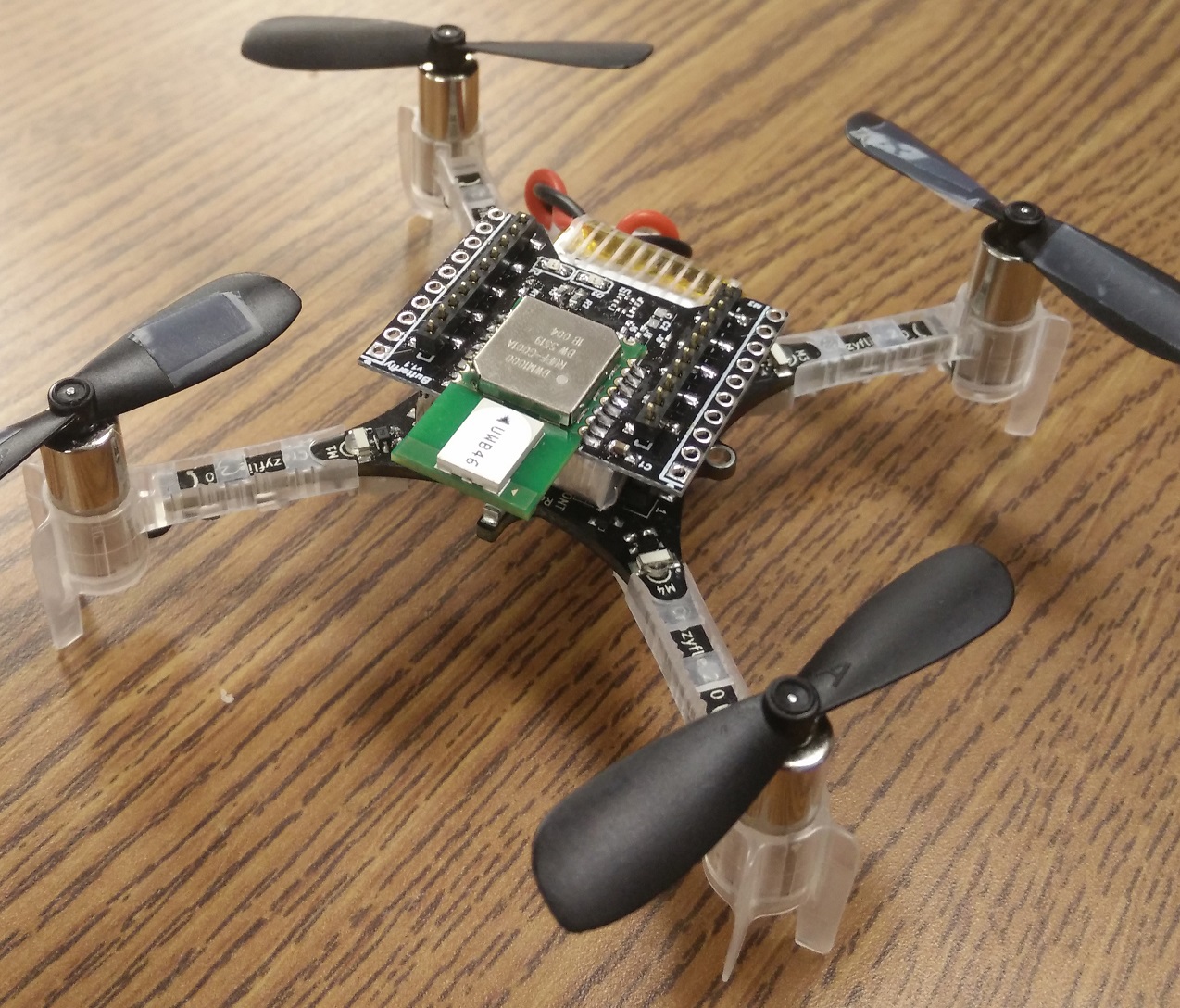}
\caption{CrazyFlie 2.0 quadrotor helicopter with UWB expansion.}
\label{fig:quad}
\end{minipage}
\vspace{-7mm}
\end{figure*}

\subsection{Experimental Setup}
We evaluate the performance of Algorithm $1$ on a custom ultra-wideband (UWB) RF testbed based on the DecaWave DW1000 IR-UWB radio\footnote{Decawave DW1000 \url{http://www.decawave.com/products/dw1000}}. The overall setup is presented in Figure \ref{fig:full_sys}. The main components of the considered testbed can be summarized as follows:

\begin{itemize}

\item The motion capture system with eight cameras capable of performing 3D rigid body position measurement with a accuracy of less than $0.5$ mm accuracy. %The system consists of an eight-camera which are deployed to provide accurate ground truth position measurements. The ground truth positions from the motion capture cameras are sent to a centralized server that uses the Robot Operating System (ROS) \cite{ros} with a custom package. 
The presented results consider the motion capture estimates as the true position, even though we qualify here that all results are accurate with respect to the motion capture accuracy. %We adopt a right-handed coordinate system where $y$ is the vertical axis, and $x$ and $z$ make up the horizontal plane. 

\item Fixed nodes consist of custom-built circuit boards equipped with ARM Cortex M4 processors of $196$ MHz powered over Ethernet and communicating by Decawave DW1000 ultra-wideband radios, as shown in Figures \ref{fig:ntbanon} and \ref{fig:ntbceil}. Each anchor performs the single and double-sided two-way range measurement with its neighbors. The used Decawave radio is equipped with a temperature-compensated crystal oscillator with frequency of $38.4$ MHz and stated frequency stability of $\pm 2$ ppm. We installed eight UWB anchor nodes in different positions in a $10\times 9$ $m^2$ laboratory. More specifically, six anchors are placed on the ceiling at about $2.5$ m hight, and the other two are spotted at the waist height of about $1$ m to disambiguate positions on the vertical axis in a better manner. %Each anchor node is fully controllable over a TCP/IP command structure from the central server. %These nodes are placed to remain mostly free from obstructions, maximizing line-of-sight barring pedestrian interference. 

\item The battery-powered mobile node is a modified CrazyFlie 2.0 helicopter\footnote{Bitcraze CrazyFlie 2.0. \url{https://www.bitcraze.io/}} is equipped with the same DW1000 radio and ARM Cortex M4 processor, as shown in Figure \ref{fig:quad}. %This allows for compatibility in the single and double-sided ranging technique used. 
\end{itemize}

\begin{figure*}[t!]
\centering
\includegraphics[width = 0.8\textwidth]{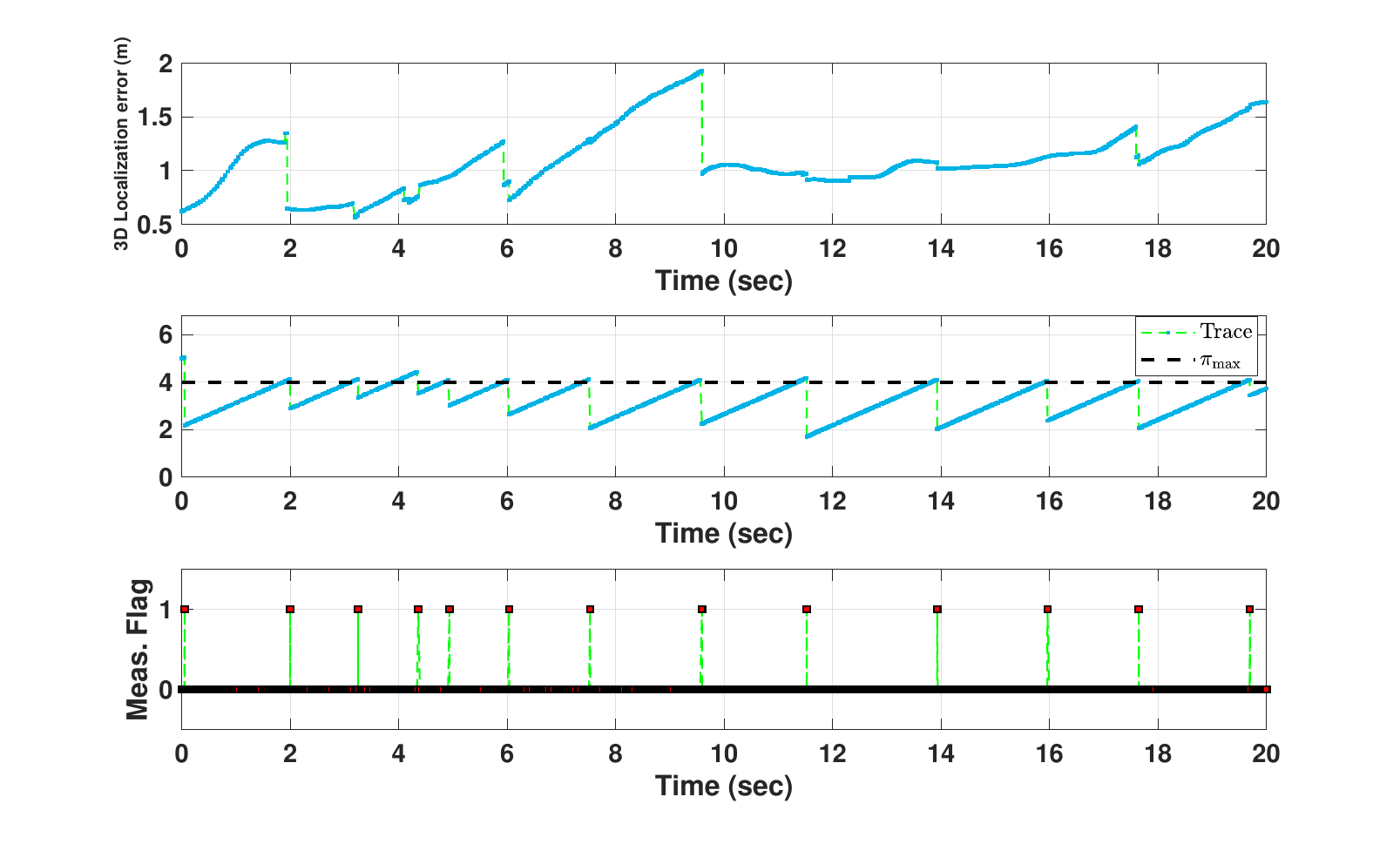}
\vspace*{-5mm}
\caption{Snapshot of $20$ seconds of the conducted experiments. The threshold is set to $4$ $m^2$. The 3D localization error and trace value $\tr(WP_{L,i|i}W^*)$ are shown in the first and second sub-figures, respectively. The measurement and diffusion flags are the same and shown in the third sub-figure in which a value of $1$ indicates of the step execution, while $0$ means skipping the step at the corresponding time instance. Time update step is executed all the time.}
\label{fig:pdkal}
\vspace*{-5mm}
\end{figure*}

\subsection{Experiments}

In the conducted experiments, we mainly consider the communication overhead and the associated accuracy on different network topologies to demonstrate the effectiveness of the proposed algorithm.  %Then, we are going to do case studies to have a satisfying evaluation. 
%We demonstrate the performance of our algorithm on distributed simultaneous localization and time synchronization problem as mentioned before. The eight static nodes are placed in distinct locations around the $10 \times 9m^2$ area. 
We aim to apply the triggering logic based on the expected estimation error of the mobile node location. The mobile node is chosen as a leader and its diffusion error covariance matrix is $P_{L,i|i}$ in Algorithm~\ref{alg:one}. 
%In other words, we use $W$ as following:
%\begin{equation}
%      W = \begin{bmatrix}
%         1 & 0 & 0 & 0 & 0\\
%         0 & 1 & 0 & 0 & 0\\
%         0 & 0 & 1 & 0 & 0
%      \end{bmatrix}.
%\end{equation}
%At a high level, the wireless network has a relevant state that varies over time. The network needs to efficiently utilize its computation and communication resources while achieving the required performance. 

To give the intuition behind evaluation of the proposed algorithm, we present the results of running a portion of the experiments in Figure \ref{fig:pdkal}. The mobile node flies at different speeds in the testing laboratory while seeking to save computational and communication resources. The threshold $\pi_{\max}$ is set to $4$ $m^2$. The second graph in Figure \ref{fig:pdkal} presents the behavior of $\tr(WP_{L,i|i}W^{T})$. All nodes execute the time update step when $\tr(WP_{L,i|i}W^{T})$ is less than $\pi_{\max}$. Consequently, we decrease the spending in terms of measurements, message exchange, and computational costs. The effect on the estimated localization error of the mobile node can be clearly seen in the first graph of Figure \ref{fig:pdkal}. As soon as $\tr(WP_{L,i|i}W^T)$ reaches the threshold $\pi_{\max}$, all nodes are triggered to start measuring and to exchange messages to decrease $\tr(WP_{L,i|i}W^T)$ back to the allowed range. The third graph in Figure \ref{fig:pdkal} demonstrates the case when the measurement update step is executed at all nodes. For instance, we can see that the measurement update step happens at the time instances $0.1,  2.7, 4.4, 5.1, 7.0, 10.1, 12.9,$ and  $16.4,$ so that we can notice the decrease in the localization error in the first graph at the same time instances.
%\begin{equation}
%Time = \begin{bmatrix}
%         0.1 &  2.7 & 4.4 &   5.1  &  7.0  & 10.1  & 12.9 &  16.4
%\end{bmatrix}\nonumber
%\end{equation}
Similarly, the diffusion update step is executed just after the measurement update step. The time update step is executed all the time. 

The CrazyFlie flied across the laboratory over four different sessions. Each session was conducted on a different day with a different number of students in the room. The path of the CrazyFlie was a random walk for each session. We repeated the experiments while setting a different threshold, then calculated the number of shared messages between nodes and the localization error reported by the motion capture system.

\subsubsection{Case Study on a Fully-Connected Network}

We illustrate the effectiveness of the proposed triggering algorithm by outlining the amount of saving on the communication overhead and the associated localization error by applying the algorithm over a fully-connected network. %We should note that our proposed algorithm could also be used with the general distributed Kalman, where the fully connected case is more applicable.
%The total duration of the session was about 10 minutes. 
\paragraph{Communication Analysis}

\begin{figure*}[tbp]
%\vspace{-0.05cm}
    \centering
    \begin{tabular}{ p{0.47\textwidth}  p{0.47\textwidth}  }
        \resizebox{0.47\textwidth}{!}{
            \begin{subfigure}[h]{0.47\textwidth}
      \centering
        \includegraphics[width = 0.8\textwidth]{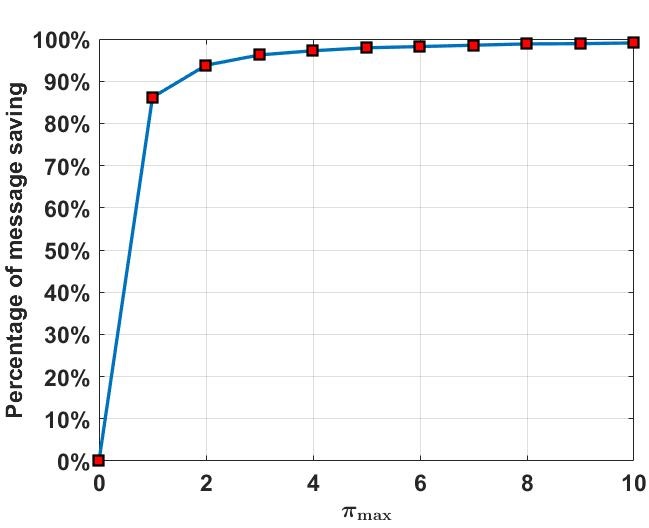}
        \caption{Effect of changing the threshold value $\pi_{\max}$ on the percentage of the saved messages for a fully connected network.}
\label{fig:thres_vs_msg}
    \end{subfigure}
       } 
   &
   \resizebox{0.47\textwidth}{!}{
            \begin{subfigure}[h]{0.47\textwidth}
      \centering
        \includegraphics[width = 0.8\textwidth]
        {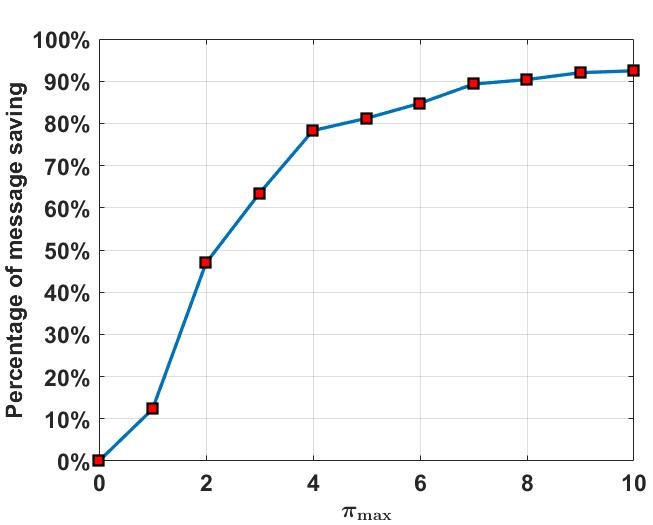}
\caption{Effect of changing the threshold value $\pi_{\max}$ on the percentage of the saved messages for a partially connected network.}
\label{fig:thres_vs_msg_notconn}
% \vspace{-1mm}
    \end{subfigure}
      }
      \\
        \resizebox{0.47\textwidth}{!}{
            \begin{subfigure}[h]{0.47\textwidth}
      \centering
        \includegraphics[width = 0.8\textwidth]
                {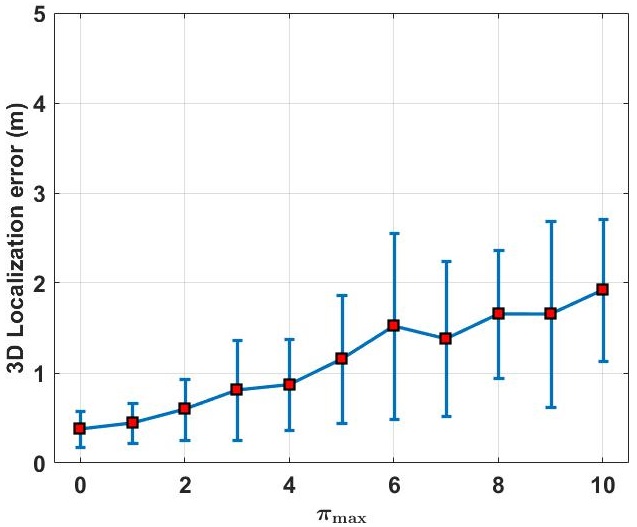}
\caption{Effect of changing the threshold value $\pi_{\max}$ on the 3D localization error of the CrazyFlie for a fully connected network.}
\label{fig:thres_vs_err_errplot}
    \end{subfigure}
       } 
   &
   \resizebox{0.47\textwidth}{!}{
            \begin{subfigure}[h]{0.47\textwidth}
      \centering
        \includegraphics[width = 0.8\textwidth]
        {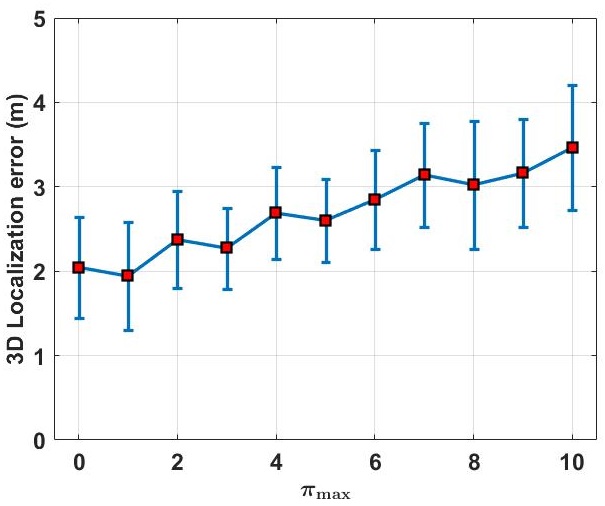}
\caption{Effect of changing the threshold value $\pi_{\max}$ on the 3D localization error of the CrazyFlie for a partially connected network.}
\label{fig:thres_vs_err_errplot_notconn}
    \end{subfigure}
      }      
  \end{tabular}
\caption{Effect of changing the threshold value $\pi_{\max}$ on a fully connected network and partially connected one.}% 
    \label{fig:disatt}%\vspace{-4mm}
\vspace{-6mm}
\end{figure*}

Figure \ref{fig:thres_vs_msg} shows the effect of changing the threshold value $\pi_{\max}$ on the percentage of the saved message for a fully-connected network. The zero threshold refers to the case of sending all messages ($1,975,632$ messages in total) and running the three steps of the algorithm in the normal mode. Setting the threshold to $1$ $m^2$ allows saving approximately $86.2\%$ of the overall number of messages. Moreover, the threshold of $5$ $m^2$ leads to saving $98\%$ of the total number of messages. %Finally, saving $99\%$ of the total number of messages can be achieved by $10m$ threshold. 
%It should be noted that every message already corresponds to the single-sided or double-sided two-way range measurements. Therefore, the reported number of messages should be multiplied by two or three, depending on the message ranging type.

\paragraph{Accuracy Analysis}

We have conducted a study to identify what is the tradeoff between the 3D-localization errors corresponding to each threshold value $\pi_{\max}$. The error plot in Figure \ref{fig:thres_vs_err_errplot} summarizes the obtained results of this case study. The red rectangles correspond to the mean values of the localization error, while the vertical lines represent the standard deviation around that mean value. At threshold $\pi_{\max} = 0$, we do not save any resources, and we achieve $0.377$ $m$ mean localization error with the standard deviation of $0.195$ $m$. While, $\pi_{\max} = 5$ $m^2$ results in a $1.155$ $m$ mean error with the standard deviation of $0.71$ $m$. Finally, $\pi_{\max} = 10$ $m^2$ allows achieving the mean error of $1.923$ $m$ with the standard deviation of $0.790$ $m$. The appropriate thresholds can be set by users based on their needs.

%{Figures/thres_vs_msg_morepts.jpg}

\begin{figure}[tb]
\centering
\includegraphics[width = 0.35\textwidth]{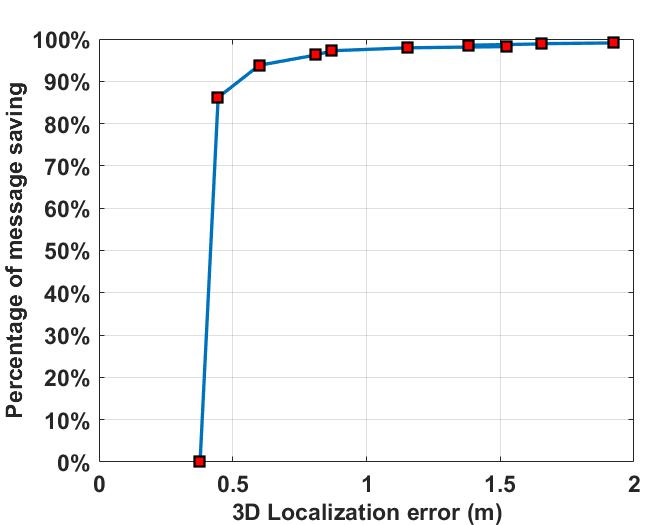}
\vspace{-2mm}
\caption{The tradeoff between the communication overhead saving and the mean 3D localization error of the CrazyFlie.}% for a fully connected network.}
\label{fig:err_vs_msg}
\vspace{-7mm}
\end{figure}
% \begin{figure}[tb]
% \centering
% \includegraphics[width = 0.45\textwidth]%{Figures/thres_vs_msg.jpg}
% {Figures/thres_vs_msg_morepts.jpg}
% \caption{Effect of changing the threshold value $\pi_{\max}$ on the percentage of the saved message for a fully connected network.}
% \label{fig:thres_vs_msg}
% \end{figure}
% 
% \begin{figure}[tb]
% \centering
% \includegraphics[width = 0.45\textwidth]{Figures/thres_vs_err_errplot.jpg}
% \caption{Effect of changing the threshold value $\pi_{\max}$ on the 3D localization error of the CrazyFlie for a fully connected network.}
% \label{fig:thres_vs_err_errplot}
% \end{figure}
% 
% 
% 
% \begin{figure}[tb]
% \centering
% \includegraphics[width = 0.45\textwidth]{Figures/thres_vs_msg_notconn.jpg}
% \caption{Effect of changing the threshold value $\pi_{\max}$ on the percentage of the saved message of the network for a partially connected network where every node is connected to only neighbors.}
% \label{fig:thres_vs_msg_notconn}
% \end{figure}
% 
% \begin{figure}[tb]
% \centering
% \includegraphics[width = 0.45\textwidth]{Figures/thres_vs_err_errplot_notconn.jpg}
% \caption{Effect of changing the threshold value $\pi_{\max}$ on the 3D localization error of the CrazyFlie for a partially connected network where every node is connected to only neighbors.}
% \label{fig:thres_vs_err_errplot_notconn}
% \end{figure}

The proposed algorithm is used to restrict the amount of processing, sensing, and communication. This restriction allows reducing dramatically the communication overhead in the network, however it may potentially lead to deterioration of network performance. We analyze the tradeoff between the number of messages sent in the wireless network and performance of estimation algorithm. Figure \ref{fig:err_vs_msg} shows the tradeoff between the communication overhead and the mean 3D localization error. Notably, saving $86.2\%$ of the communication overhead leads to the increase in the localization error by $16.57\%$. This was calculated by considering the mean localization error plus the standard deviation at thresholds $0$ and $1$.

%\vspace{-1mm}

\subsubsection{Case Study on a Partially Connected Network}
We considered another case study in which every node among the nine considered ones is connected to only four neighbors instead of eight neighbors in the previous case study. We analyze the communication saving and the associated localization error. 
%\paragraph{Communication Analysis:}
Figure \ref{fig:thres_vs_msg_notconn} summarizes the observed results. It should be noted that setting the threshold to $5$  leads to saving approximately $81.2\%$ of the overall number of messages. Moreover, the threshold of $\pi_{\max} = 9$ $m^2$ allows saving $92\%$ of the total number of messages. The error plot in Figure \ref{fig:thres_vs_err_errplot_notconn} summarizes the obtained results of this case study. Here, the red rectangles correspond to the mean values of the localization error, while the vertical lines represent the standard deviation around that mean value. At threshold $\pi_{\max} = 0$ $m^2$, we do not save any resources, and we can achieve $2$ $m$ mean localization error with standard deviation of $0.6$ $m$ in the case when the network is partially connected, as described before. While $\pi_{\max} = 5$ $m^2$ results in the mean error of $2.6$ $m$ with the standard deviation of $0.49$ $m$. %Finally, $\pi_{\max} = 10$ $m^2$ leads to the mean error of $3.46$ $m$ mean error with the standard deviation of $0.74$ $m$.
%It is up to the application need to set the appropriate threshold based on its need.

\section{Conclusion} \label{sec:conc}
%\vspace{-3mm}
In the present study, we investigated the energy-aware aspect of the distributed estimation problem with regard to the multi-sensor system with event-triggered processing schedules. More specifically, we proposed an unbiased event-triggered distributed diffusion Kalman filter for wireless networks based on the diffusion covariance matrix. We tested the proposed algorithm on the distributed localization and time synchronization application. Several experiments were conducted using the real custom ultra-wideband wireless anchor nodes and mobile quadrotor. The obtained results indicate that the proposed algorithm is has robust performance, and is efficient in terms of using computational and communication resources. %Future directions will deal with testing the algorithm with a large-scale system.
\vspace{-2mm}
\section*{Acknowledgements}
\vspace{-1mm}
We gratefully acknowledge partial financial support by the project justITSELF funded by the European Research Council (ERC) under grant agreement No 817629 and the project interACT under grant agreement No 723395; both projects are funded within the EU Horizon 2020 program.


%%%%%%%%%%%%%%%%%%%%%%%%%%%%%%%%%%%%%%%%%%%%%%%%%%%%%%%%%%%%%%%%%%%%%%%%%%%%%%%%

%%%%%%%%%%%%%%%%%%%%%%%%%%%%%%%%%%%%%%%%%%%%%%%%%%%%%%%%%%%%%%%%%%%%%%%%%%%%%%%%

%%%%%%%%%%%%%%%%%%%%%%%%%%%%%%%%%%%%%%%%%%%%%%%%%%%%%%%%%%%%%%%%%%%%%%%%%%%%%%%%
%\section*{APPENDIX}


%\section*{ACKNOWLEDGMENT}

%%%%%%%%%%%%%%%%%%%%%%%%%%%%%%%%%%%%%%%%%%%%%%%%%%%%%%%%%%%%%%%%%%%%%%%%%%%%%%%%
%title, author names, pages, year, conference/journal/book name, and optionally volume and nr (for journals).
% no abbreviation ACC, CDC ..
%Proceedings not Proc.
\bibliographystyle{IEEEtran}
\bibliography{ref}

%\clearpage
%\clearpage
%\input{sections/appendix.tex}
%\begin{thebibliography}{99}

%\bibitem{c1} G. O. Young, ÒSynthetic structure of industrial plastics (Book style with paper title and editor),Ó 	in Plastics, 2nd ed. vol. 3, J. Peters, Ed.  New York: McGraw-Hill, 1964, pp. 15Ð64.

%\end{thebibliography}

\end{document}